\documentclass[journal]{IEEEtran}
\usepackage{ifpdf}
 \ifpdf
 \else
 \fi

\ifCLASSINFOpdf
  \usepackage[pdftex]{graphicx}
  \graphicspath{./}
\else
  \usepackage[pdftex]{graphicx}
  \graphicspath{{./}}
\fi
\begin{document}
%
\title{A Control System of New Magnet Power Converter for J-PARC Main Ring Upgrade}
%
%
%

\author{
Tetsushi~Shimogawa$^{*}$, 
Yoshinori~Kurimoto,
Yuichi~Morita,
Kazuki~Miura and 
Daichi~Naito
\thanks{The authors are with the Magnet and Power Converter Group, Accelerator Laboratory, High Energy Accelerator Research Organization, Department of Acelerator Science, Graduate University for Advanced Studies (SOKENDAI), Tokai-mura, Ibaraki, Japan.}
\thanks{* T.~Shimogawa is the corresponding author. Email: tetsus@post.j-parc.jp}
}
\maketitle

\begin{abstract}
  Japan Proton Accelerator Research Complex (J-PARC) aims at a MW-class proton accelerator facility. One of the promising solutions for increasing the beam power of the Main Ring (MR) is converters of the main magnets with new ones for this upgrade. We have a plan to replace and develop the power converters of main magnets for this upgrade. According to develop the new power converter. We have developed a new control system for the new power converters. This control system consists of four parts : the current measuring device, the feedback control system, the gate pulse generator and the slow control system. Considering a reproducibility in the mass-production and the facilitation of the control algorithm, the digital control system is adopted. We will report the design of this control system and some test result with a new power converter for small quadrupole magnets.
\end{abstract}


%
\IEEEpeerreviewmaketitle

%
%
%
%
\section{Introduction}
\begin{figure}[!t]
  \centering
  \includegraphics[width=7cm]{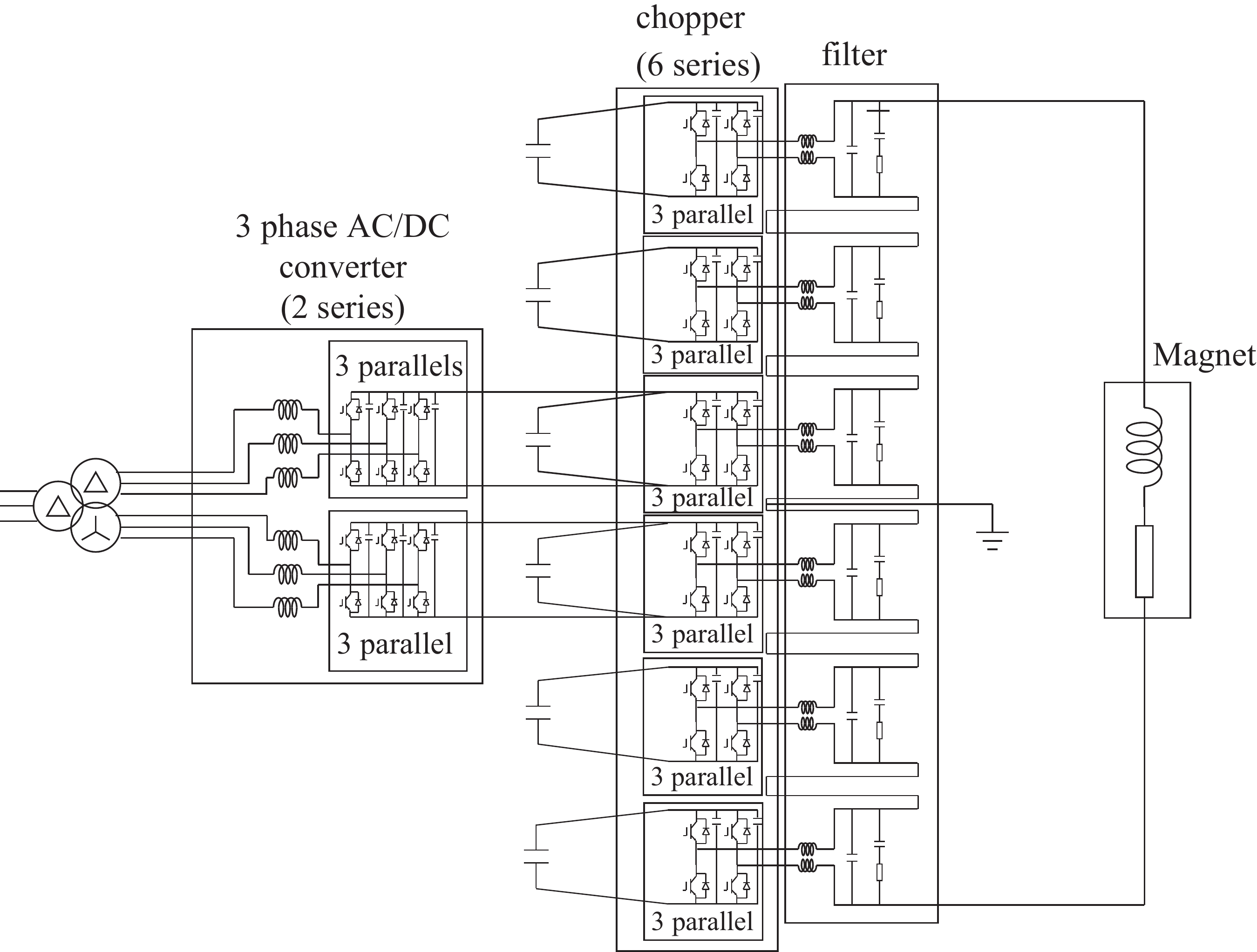}
  \includegraphics[width=7cm]{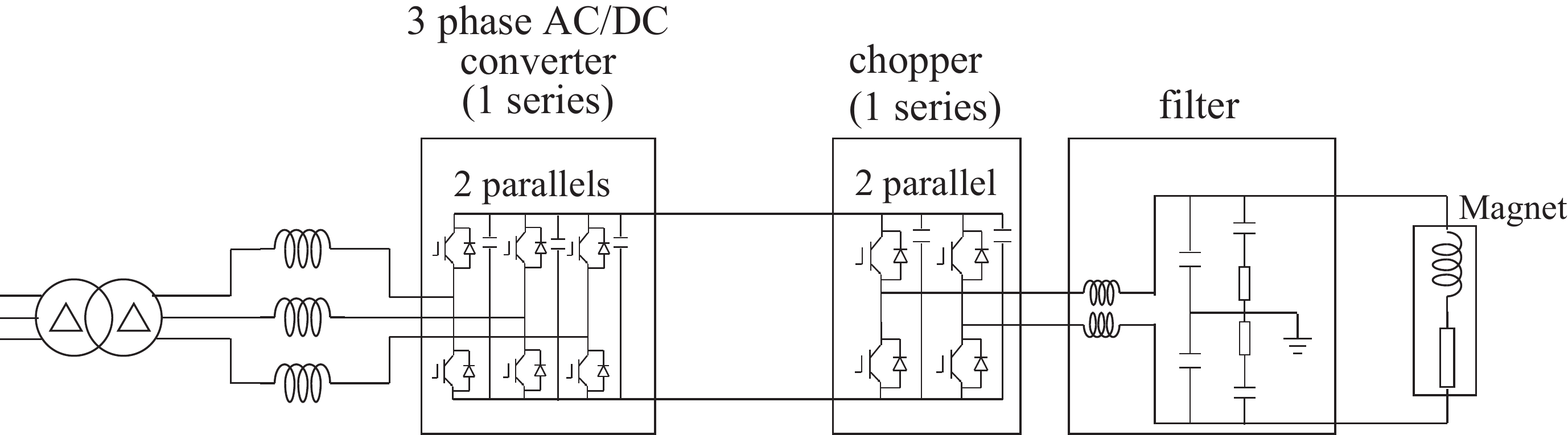}
  \caption{Schematics of the new power converters for bending magnets (top) and small quadrupole magnets (bottom).}
  \label{fig:PS_schematic}
\end{figure}

\IEEEPARstart{J}{-PARC} MR is a high intensity proton synchrotron for the particle and nuclear physics. 
The proton beam is accelerated from 3 GeV to 30 GeV in the MR and provided to experimental facilities\cite{J-PARCMR}. 
The intensity of the extraction beam has reached 500 kW.
However, We still need to increase the beam intenisity to over a MW to maintain the international competitiveness.
One of the promising solutions for increasing the beam power is to shorten the repetition period from current rating 2.48 sec to 1.3 sec.
For shorter repetition period, the driving current of electromagnets must be ramped faster with higher voltage power converters. 
Consequently, we have planned to develop new power converters for the main magnets, such as bending, quadrupole and sextupole magnets, and replace the present power converters with the new ones. \\
We need the power converters shown in Table~\ref{tab:PS_list}. Several identical power converters are also needed. 
For these requirements, we design new power converters which consist of our developed power units. This power unit is a IGBT (Insulated Gate Bipolar Transistors) half-bridge circuit at 1700 V and 525 A for the rated voltage and current, respectively. The present design of new power converters for bending magnets and small quadrupole magnets are shown in Figure~\ref{fig:PS_schematic}.
For example, new power converter for the bending magnets have two series of AC/DC converters and 6 series of choppers. In addition, three AC/DC converters and three choppers are connected in parallel. On the other hand, new power converter for the small quadrupole magnets consists of only two parallels of AC/DC converters and choppers, which are not connected in series.
Both of the AC/DC converter and the chopper require a control system which satisfies following requirements.

\begin{itemize}
\item{Summarizing alarms and failure protection}
\item{Real time feedback control of a power converter with several monitors}
\item{Generating gate pulses for switching devices in the power units}
\item{Managing sequence of power converter operation}
\item{Monitoring status from host system via Ethernet}
\item{Applicable to various configurations of the power converters}
\end{itemize} 

We have developed the control system based on these requirements.

\begin{table}[!t]
\renewcommand{\arraystretch}{1.3}
\caption{List of current and voltage rated for new power converters.}
\label{tab:PS_list}
\centering
\begin{tabular}{|c||c|c|c|c|}
  \hline
  & Flat & Flat & Output Voltage & Number  \\
  Magnet & Bottom & Top & at 1.3 sec & of \\
  family & Current & Current & repetition & power \\
   & [A] & Flat [A] & [kV] & converters \\
\hline
Bending & 190 & 1570 & 6.0 & 6 \\
\hline
Large quadrupole  & 80 & 1000 & 7.0 & 4 \\
\hline
Small quadrupole & 70 & 1000 & 1.5 & 7 \\
\hline
Sextupole & 20 & 200 & 0.8 & 3 \\
\hline
\end{tabular}
\end{table}

\section{Hardware design}
\label{sec:HardwareCont}
\begin{figure*}[!t]
  \centering
  \includegraphics[width=13cm]{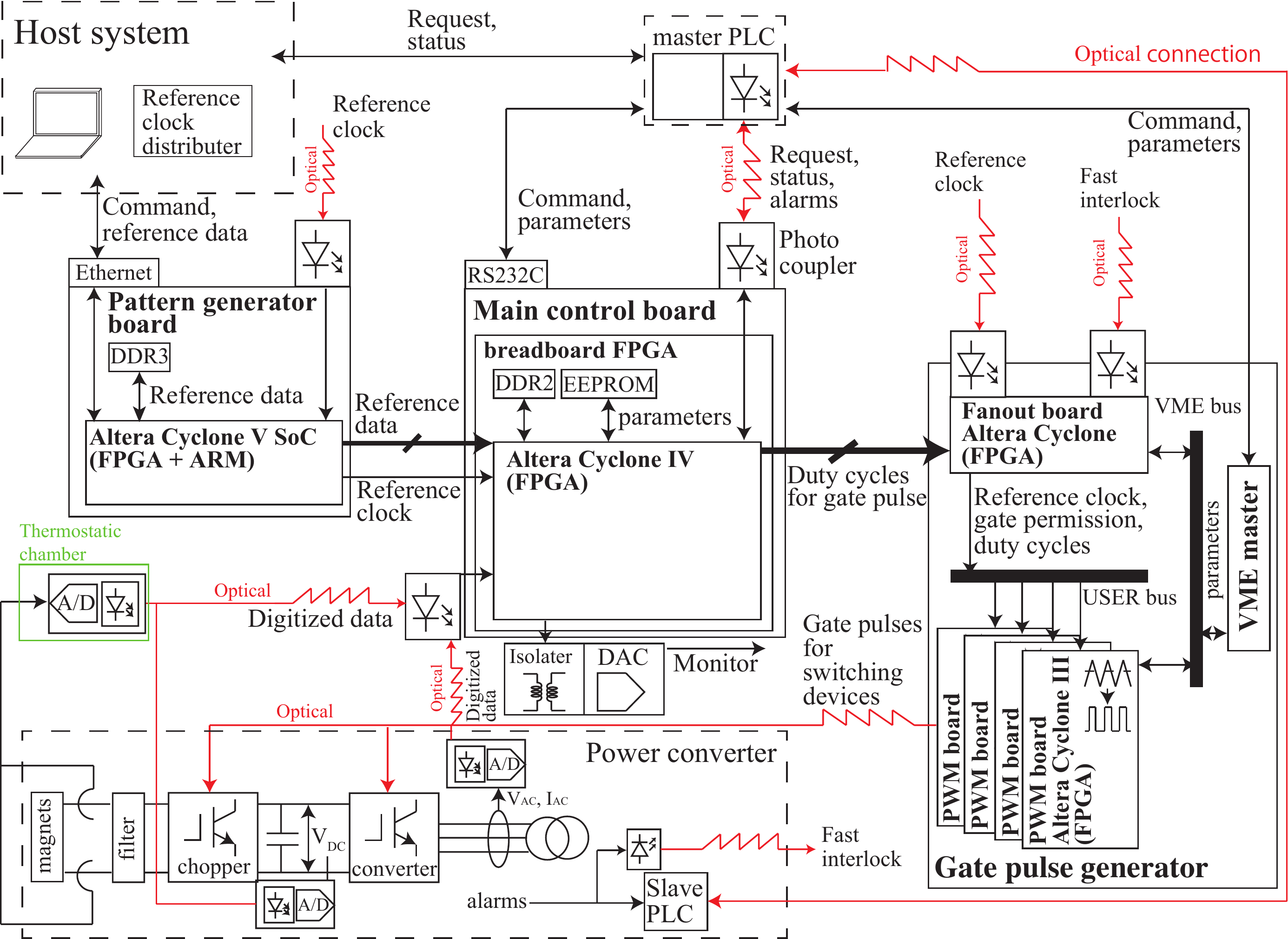}
  \caption{Overview of the controller.}
  \label{fig:overview_cont}
\end{figure*}

The controller consists of four main parts: the current measuring device, the feedback control system, the gate pulse generator for the IGBT in the power units and the slow control system, whose rules are summarizing alarms, managing sequence and failure protection, as shown in Figure~\ref{fig:overview_cont}. Analog components are included only in the current measuring device while the other three parts are purely digital. \\

At the first stage of the analog processing, the DCCT transforms the large output current into the small one at a ratio. The 1 $\Omega$ burden resister, which is on our developed 24-bits analog-to-digital converter board, changes the current signal to the voltage before the digitization. The details of our analog-to-digital converter board are described in \cite{MRMAG2014precise}. For precise current regulation, these analog devices are electrically isolated from the digital parts. These details are also described in our previous report\cite{MRMAG2014precise}. In additions, our precise current regulation must be maintained for a long time because not a few accelerators are required to be continuously operated for several months. In such cases, the temperature-dependent variation in the burden resistance and the gain drift of the analog-to-digital converter must be taken into account. For this purpose, we developed the thermostatic chamber, which will be described in Section~\ref{sec:TEMPUNIT}. \\
Comparing the digitized output current with the reference, the feedback system calculates the duty cycles of the IGBT switches. The calculated duty cycles are converted to the IGBT gate pulses using the gate pulse generator. For both digital systems, we adopted the full FPGA-based design. In fact, many accelerators utilize several types of electrode magnets which require different power converters for their drivers in term of their required current and voltage. Therefor, we must introduce several power converters, where the number of IGBT units connected in series or parallel are different. The FPGA-based system enables us to flexibly adapt the controllers to such different power converters. The details of these digital hardware will be described in Section~\ref{sec:FullFPGA}.

\subsection{Temperature Control of the analog-to-digital converter}
\label{sec:TEMPUNIT}
The long-term stability of the current regulation is mainly affected by the thermal dependence of the burden resister and analog-to-digital conversion, whose thermal coefficients are 2.5 ppm/$^{\circ}\mathrm{C}$ and 2.0 ppm/$^{\circ}\mathrm{C}$ should be controlled within $\pm$ 2.0-2.5 $^{\circ}\mathrm{C}$ to achieve at the stability of measured current less than 10 ppm. In addition, the temperature around the peripheral ICs on the analog-to-digital converter board should be also controlled to secure the stable output current measurement. Our developed thermostatic chamber consists of a thermally insulated chamber accommodating the analog-to-digital converter board, a thermoelectric device resistance thermometer (PT100) for measuring the temperature inside the thermostat chamber.
Since the energy dissipation of the burden resister is the main source of heat, the Peltier module and PT100 are located just below the burden resister.

\begin{figure}[!t]
  \centering
  \includegraphics[width=7cm]{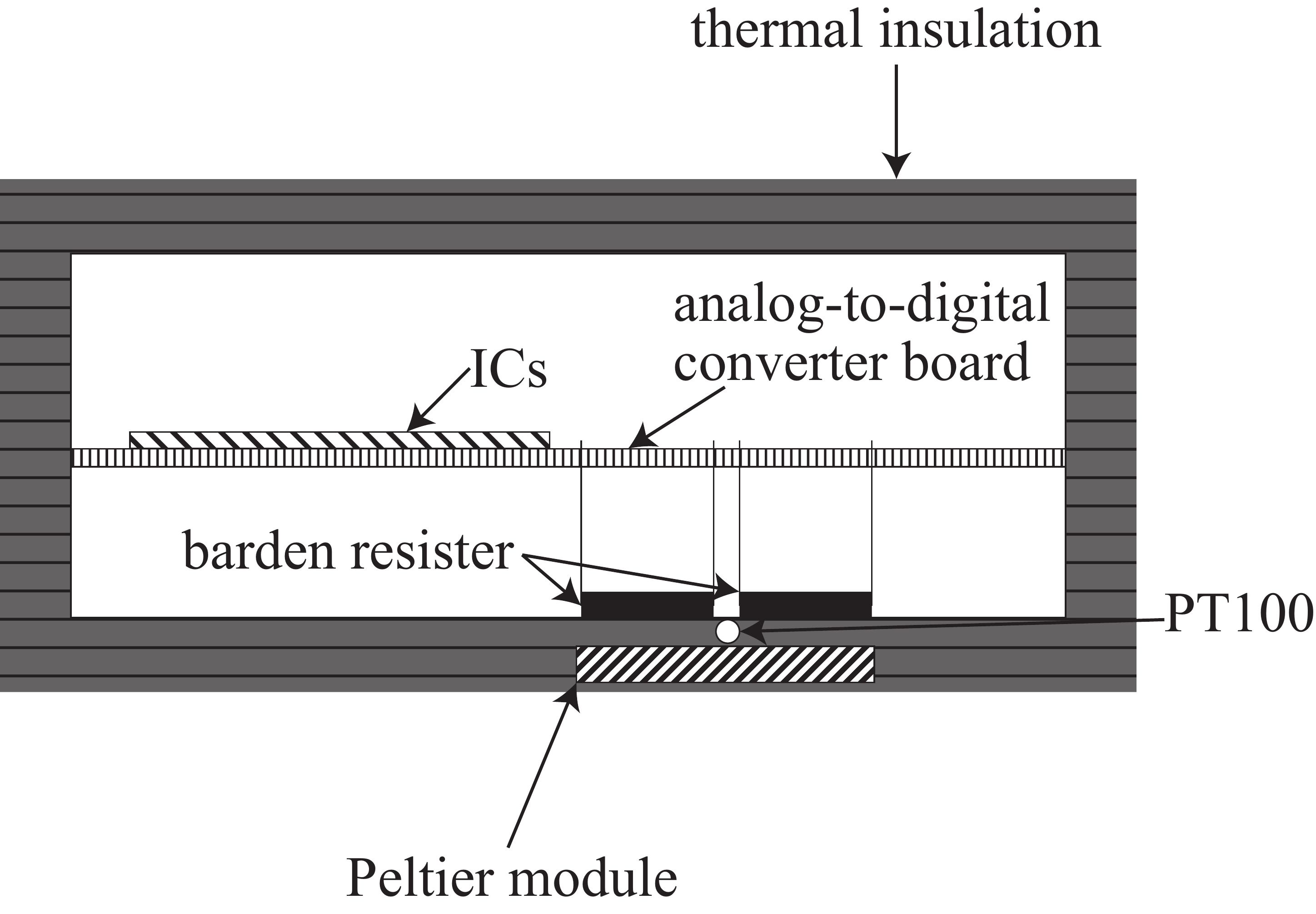}
  \caption{Illustration of the thermostatic chamber.}
  \label{fig:thermostatic_chamber}
\end{figure}

\begin{figure}[!t]
  \centering
  \includegraphics[width=7cm]{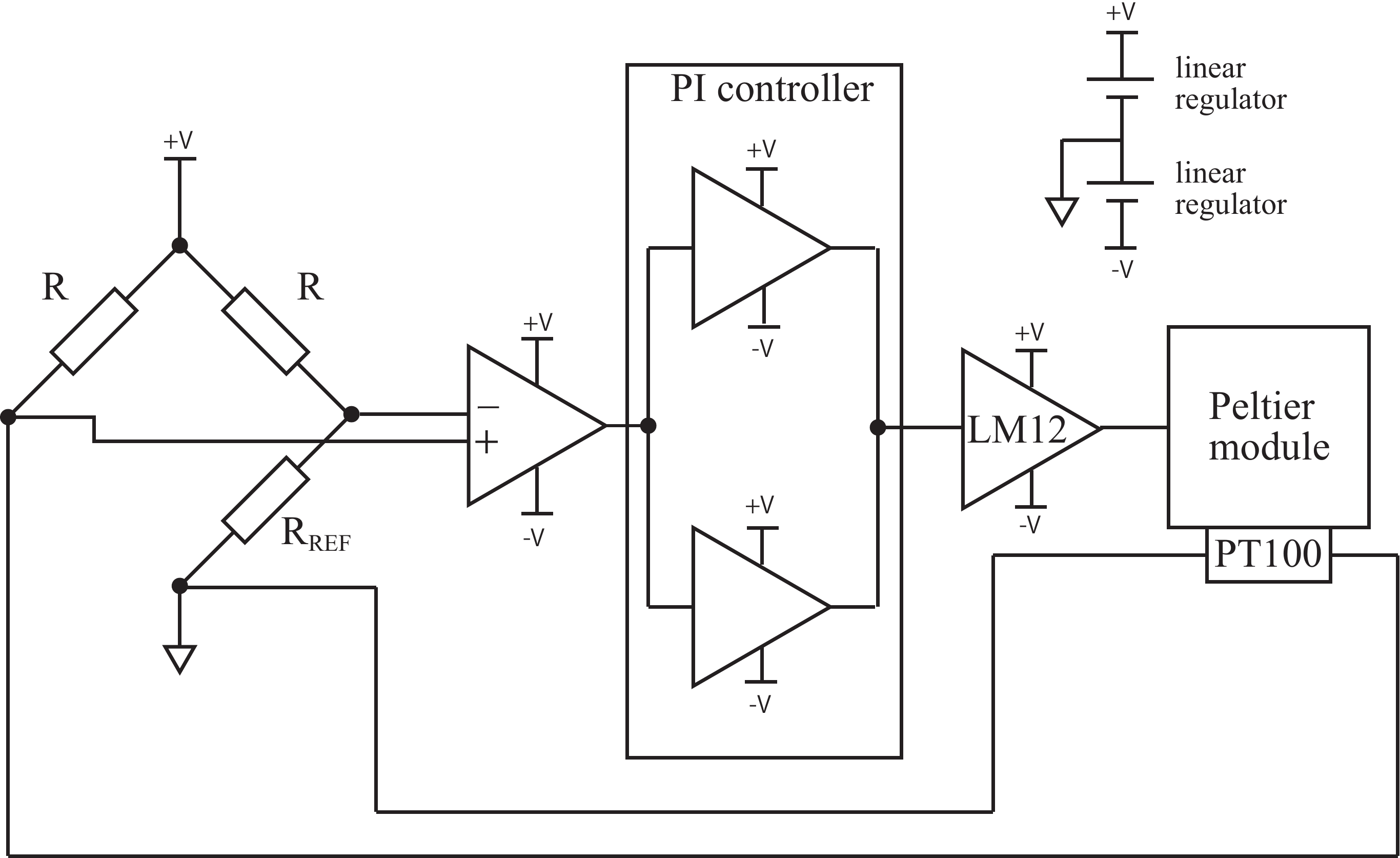}
  \caption{Schematic view of the thermostatic chamber controller.}
  \label{fig:control_circuit_temp}
\end{figure}

In addition to the thermostatic chamber, we developed a Peltier controller only with analog devices as shown in Figure~\ref{fig:control_circuit_temp}. This is because the large and fast components of digital signals may adversely affect the current measurement in the chamber. For the same reason, we adopted linear DC regulators in stead of switching regulators. In the controller, the Peltier module is driven by a power operational amplifier (LM12) through a proportional-integral (PI) controller which is designed with instrumentation amplifiers. The input of the PI controller is the measured temperature deviation from the reference which is obtained by the Wheatstone bridge.
The reference temperature is determined by the R$_{REF}$.
The reference temperature can be adjust with the variable resister (RV).
We actually checked the performance of the chamber by changing the environmental temperature from 30 to 40 $^{\circ}\mathrm{C}$.
As a result, the temperatures of the burden resister and internal air were controlled within 2.0 and 2.5 $^{\circ}\mathrm{C}$, respectively.

\subsection{Full FPGA-based Controller}
\label{sec:FullFPGA}
\begin{figure}[!t]
 \centering
 \includegraphics[width=8cm]{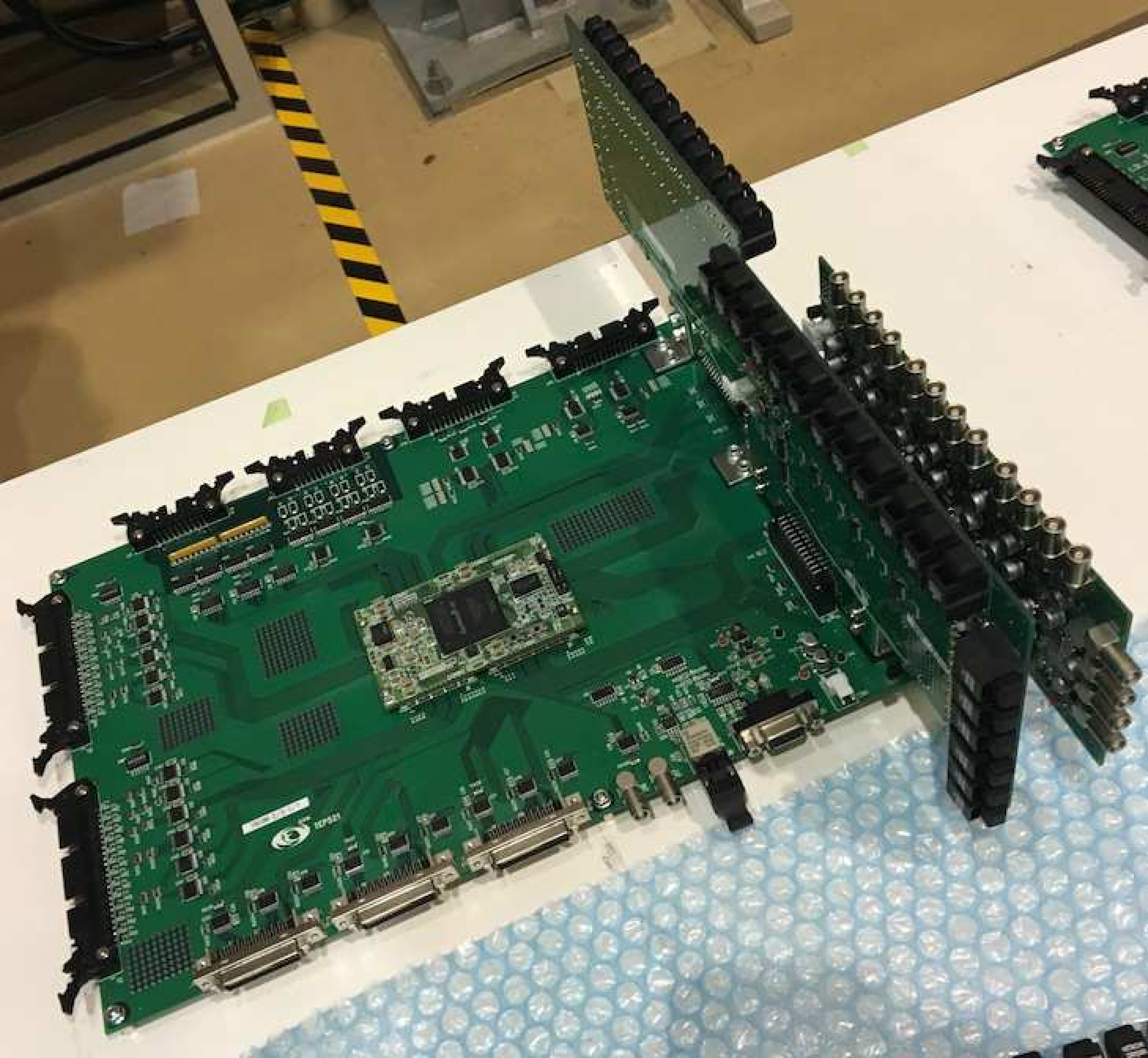}
 \caption{Picture of main control board.}
 \label{fig:CTRL_board}
\end{figure}

Two digital boards, which are the pattern generator board and the main control board in Figure~\ref{fig:overview_cont}, play main roles of the feedback system. As we will describe in Section~\ref{sec:Test}, our power converter must regulate current with predetermined pulse shapes, which are iteratively modified to precisely match the particle acceleration. Therefore, the reference current and voltage for the feedback system are neither constants nor simple time-dependent functions but rather arbitrary patterns". The pattern generator board\cite{SoC} provides the digital references of the current and voltage to the main control board using their 20 bit parallel interface.  The updates of these reference are performed at a frequency of 96 kHz in real time. For instance, a pulse shape with one second period requires 96 kilo-words for one pattern. The reference patterns are stored in double-data-rate3 synchronous dynamic RAM (DDR3). \\
The main control board performs real-time control to provide the duty cycles using the digitized measured current, the reference patterns and parameters such as feedback gains. The control algorithm will be described in Section~\ref{sec:Soft}. 
The main part of this board is a breadboard on which a FPGA (Altera Cyclone IV E), double-data-rate2 synchronous dynamic RAM (DDR2) and serial FLASH-ROM are mounted.
The firmware for the AC/DC converters and choppers as described in Section~\ref{sec:Soft} is implemented in this FPGA. 
In addition, This main control board has a lot of general IO ports connecting FPGA breadboard to realize these specification as bellow. 
\begin{itemize}
	\item{Communication to the touch panel (RS232C port): modifying parameters for control}
	\item{Connection of PLC directory (Photo-coupler contactor IO) : Receiving and sending requests and status.}
	\item{Interface for the external digital boards (GPIO port) : Deceiving and sending the reference data, the duty cycles of gate pulses and status.}
	\item{Flexible extend bus : Connecting our developed daughter boards for receiving and sending high frequency signals such as analog-to-digital converters and digital-to-analog converters as shown in Figure~\ref{fig:CTRL_board}. The digital interfaces to all daughter boards are electrically isolated using databus isolators so that noise transferred between different boards can be minimized. }
\end{itemize}

\begin{figure}[!t]
 \centering
 \includegraphics[width=3cm]{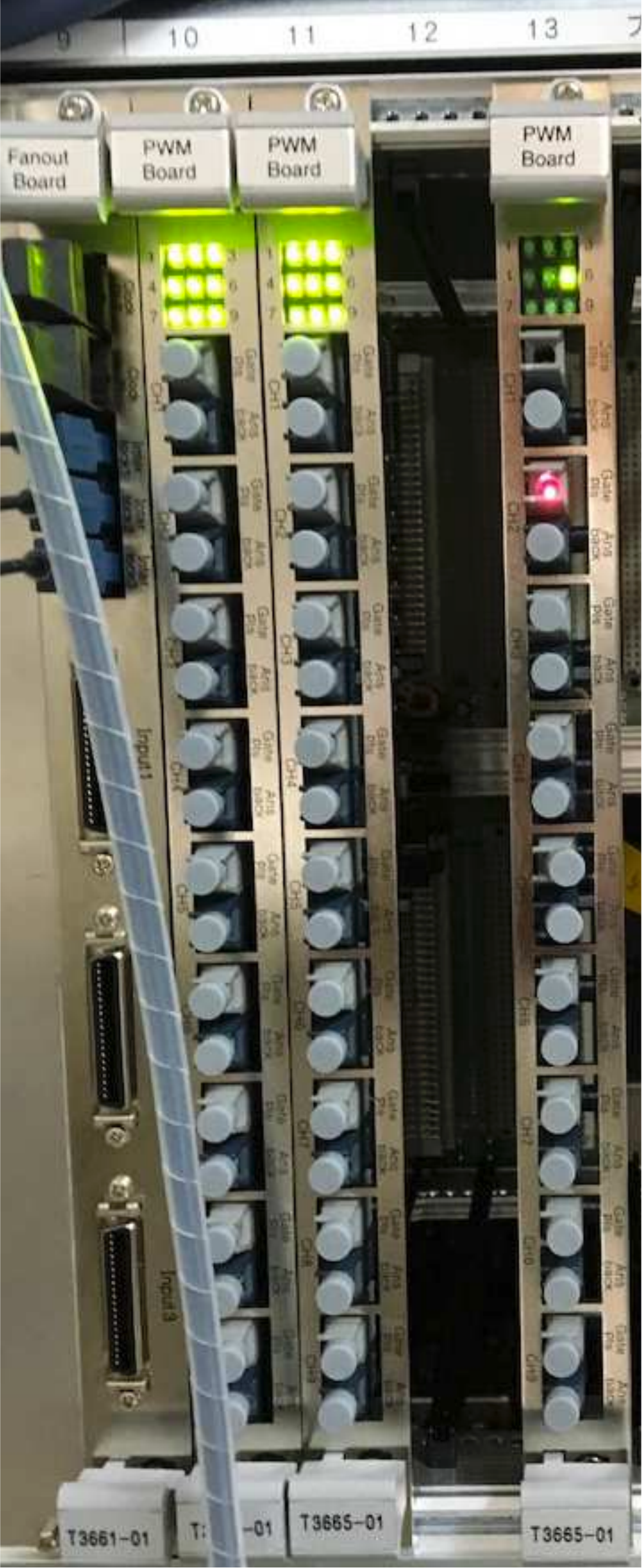}
 \caption{Picture of gate pulse generator based on VME. The Fanout board is left-side of VME crate and the others are PWM boards.}
 \label{fig:PWM_unit}
\end{figure}

The 16-bit duty cycles, which are provided by the control boards, are transferred to the gate pulse generator. This has been designed based on VME (VERSAmodule Eurocard). The gate pulse generator, shown in Figure~\ref{fig:PWM_unit}, consists of a VME crate, a board for receiving duty cycles from control board (Fanout board) and multiple Pulse Width Modulation (PWM) boards which generate the gate pulses. These boards are inserted into the VME crate and communicate with each other through data buses on the backplane of the crate. There are two type of data buses called VME bus and USER bus. The parameters of these boards, such as period of triangular waves used in the PWM boards, are modified via VME bus. The duty cycles are distributed to the PWM boards through the Fanout board using USER bus. With modifying the number of PWM boards, the number of gate signals  depending on the configuration of the power converter can be adjusted. In addition, this gate generator module has emergency gate closing function, called fast interlock system, by the direct detection of fast interlock signals from the power circuit. This fast interlock signal is a integrated signal of alarms such as over-current and over-voltage, which require immediate shutdowns of the power converter. This enables the controller to turn off all IGBTs within 10 $\mu$s after such emergency events.

\subsection{Slow control system}
These alarms are also independently detected the digital input modules of a commercially available programmable logic controller (PLC, YOKOGAWA FA-M3 series). This PLC is located in the cubicle and acts as a slave with an optical link. The master PLC is integrated in the controller so that the controller can inform users of the individual alarms. The digital input modules of the slave also detects other alarms which do not require a immediate shutdown such as over-temperature alarms. These alarms are handled as software alarm by the master PLC.  

\section{Software Design of the Controller}
\label{sec:Soft}
In this section, the control algorithms for the AC/DC converter and chopper are separately 
described.
\subsection{AC/DC Converter}
\label{sec:ACDCConv}
\subsubsection{Control of the DC Voltage and Power factor}
\label{sec:DCVPF}
\begin{figure}[!t]
\centering
\includegraphics[width=8cm]{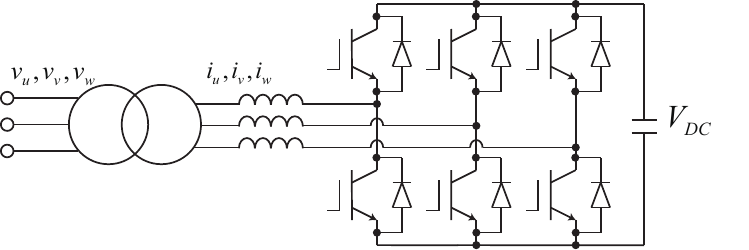}
\caption{The measurements required by the controller for the AC/DC converter }
\label{fig:ConvMeas}
\end{figure}
\begin{figure}[!t]
\centering
\includegraphics[width=8cm]{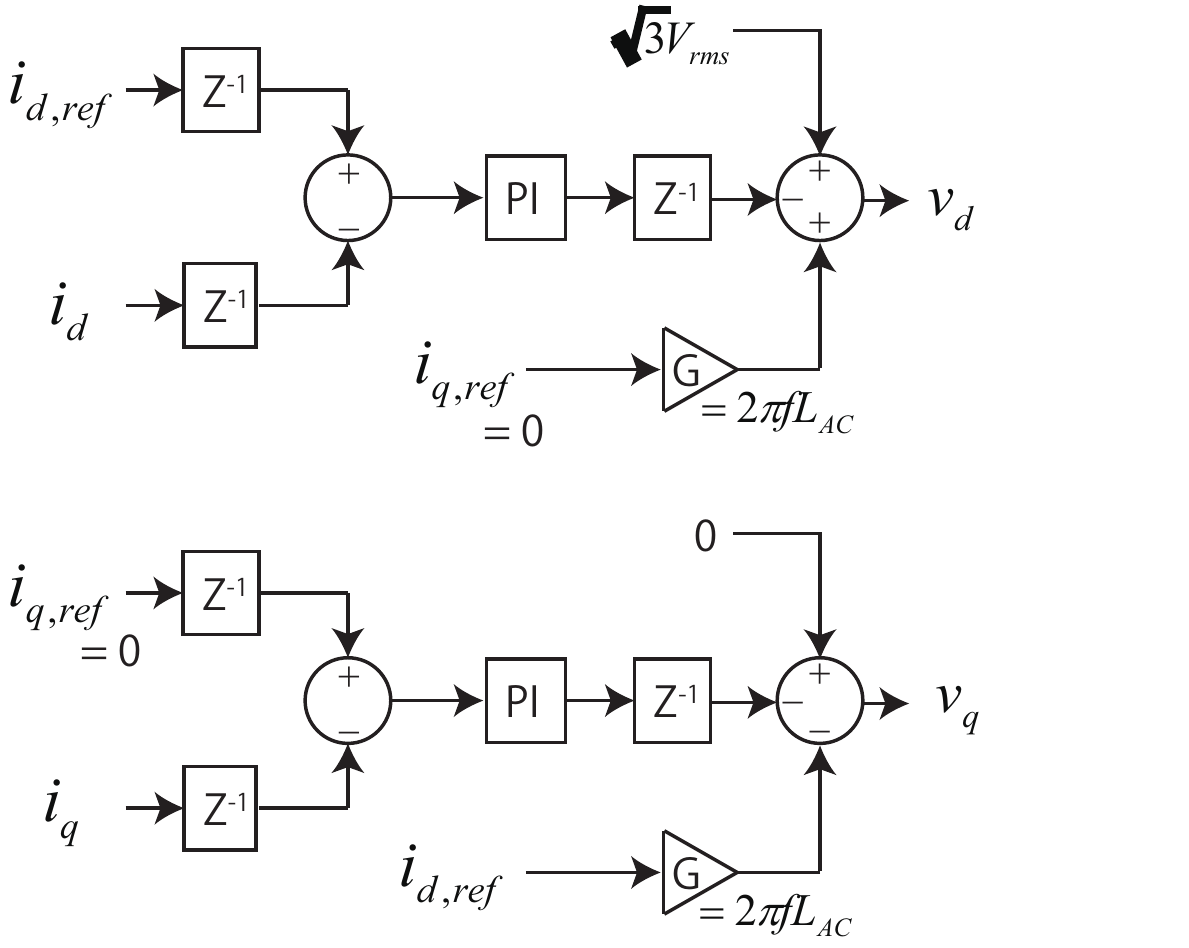}
\caption{The block diagram of the power factor control.}
\label{fig:PFactCont}
\end{figure}

The control algorithm of the AC/DC Converter has been designed based on the mathematical model reported in \cite{blasko1997new}.  
This requires seven measurements, which are the DC voltage ($V_{DC}$), three phase AC voltages ($v_{u}, v_{v}, v_{w}$) and currents ($i_{u}, i_{v}, i_{w}$) shown in Figure~\ref{fig:ConvMeas}.
\begin{figure}[!t]
\centering
\includegraphics[width=8cm]{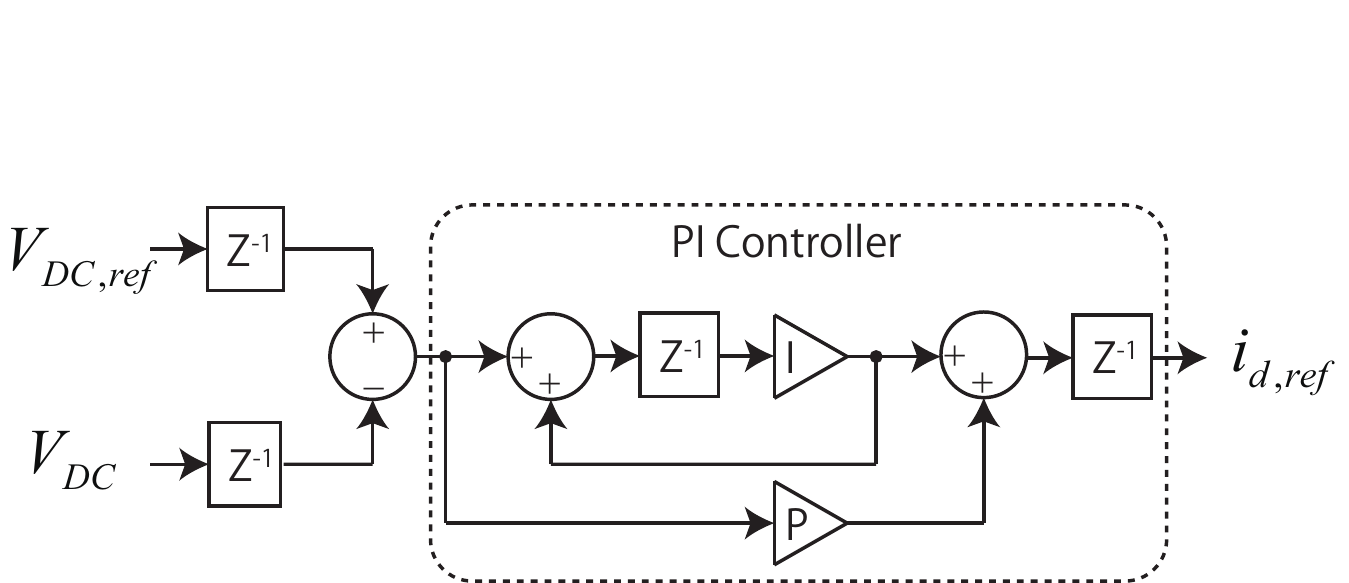}
\caption{The block diagram of the DC voltage control. In later figures, a PI controller is referred as "PI".}
\label{fig:VdcCont}
\end{figure}   
Figure~\ref{fig:VdcCont} shows the block diagram of the DC voltage feedback. The proportional-integral (PI) operation on the deviation of the DC voltage from its reference ($V_{DC,ref}$) is obtained as the reference of $i_d$ ($i_{d,ref}$). The $i_d$ is the $d$ component of the AC input current. The measured $i_d$ and $i_q$ are 
obtained as 
\begin{eqnarray}
\label{eq:abcdq}
\left( \begin{array}{c} i_{d}\\ i_{q}\\ \end{array} \right) = \nonumber \\ & \hspace{-15mm} \sqrt{\frac{2}{3}}\left(\begin{array}{ccc} \sin\phi \! & \sin(\phi-\frac{2\pi}{3}) \! & \sin(\phi-\frac{4\pi}{3}) \! \\ -\cos\phi \! & -\cos(\phi-\frac{2\pi}{3}) \! & -\cos(\phi-\frac{4\pi}{3}) \! \\  \end{array} \right) \nonumber \\
&  \times \! \left( \begin{array}{c} i_{u} \\ i_{v}\\ i_{w} \\ \end{array} \right)
\end{eqnarray}
, where $i_u$, $i_v$ and $i_w$ are three-phase currents indicated in Figure~\ref{fig:ConvMeas} and $\phi$ is the measured phase of the main grid obtained in the way described in Section~\ref{sec:GridSync}. The measured $i_d$ and $i_q$ are used for the power factor control as shown in Figure~\ref{fig:PFactCont}.  In this case, the reference of the $i_q$ indicated as $i_{q,ref}$ in the figure is set to zero so that the power factor is controlled to be unity. The obtained $d-q$ voltages indicated as $v_d$ and $v_q$ are converted 
into the three-phase voltages as 
\begin{eqnarray}
\label{eq:dqabc}
\left( \begin{array}{c} v_{u}\\ v_{v}\\ v_{w} \end{array} \right) = & \sqrt{\frac{2}{3}}\left(\begin{array}{cc} \sin\phi & -\cos\phi \\ \sin(\phi-\frac{2\pi}{3}) & -\cos(\phi-\frac{2\pi}{3}) \\ \sin(\phi-\frac{4\pi}{3})  & -\cos(\phi-\frac{4\pi}{3}) \\  \end{array} \right) \nonumber \\ 
& \times \left( \begin{array}{c} v_{d} \\ v_{q} \\ \end{array} \right).
\end{eqnarray}
The $v_u$, $v_v$ and $v_w$ in Equation~\ref{eq:dqabc} are normalized by the DC voltage and used for the duties for the IGBT gate pulses of the AC/DC converter. The $f$, $L_{AC}$ and $V_{rms}$ mean the grid frequency [Hz], the inductance [H] of the AC reactor and the rms voltage [V] of the secondary winding of the power transformer, respectively. A $Z^{-1}$ in Figure~\ref{fig:VdcCont} and \ref{fig:PFactCont} means a delay by one sample period. The sampling period for the feedback control of the AC/DC converter is 1 $\mu$s unless otherwise specified.
 
\subsubsection{Grid Synchronization}
\label{sec:GridSync}
\begin{figure}[!t]
\centering
\includegraphics[width=8.5cm]{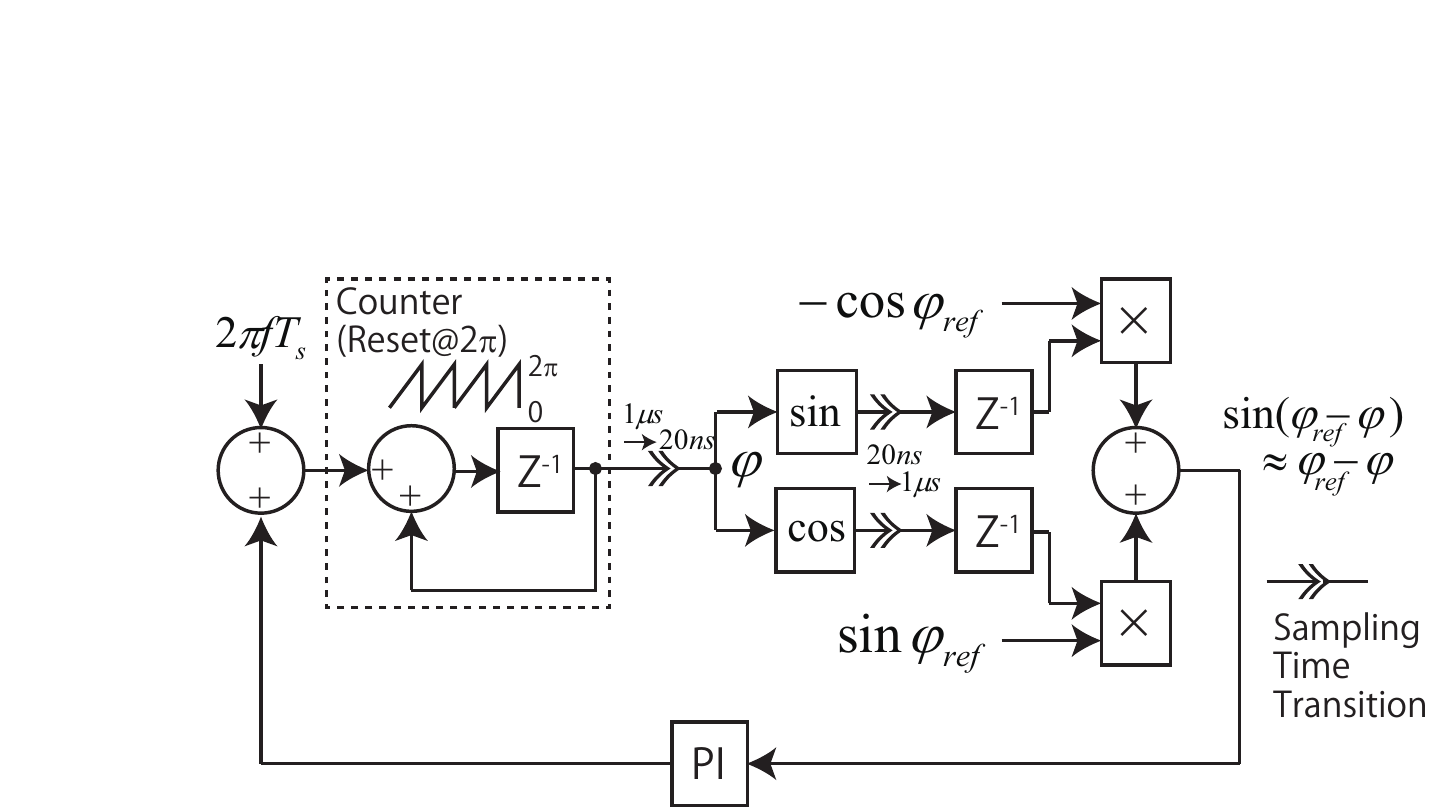}
\caption{The block diagram for measuring the phase of the main grid.}
\label{fig:GridSync}
\end{figure}  
The control of the AC/DC converter described in Section~\ref{sec:DCVPF} involves  measuring the phase of the main grid indicated as $\phi$ in Equation~\ref{eq:abcdq} and \ref{eq:dqabc}.  For this purpose, we have adopted the PLL (Phase Lock Loop) technique reviewed in \cite{blaabjerg2006overview}  and developed the Z-domain (discrete-time) algorithm so that the PLL function can be implemented in FPGA. Figure~\ref{fig:GridSync} shows the block diagram of our developed Z-domain PLL. The $\sin\phi_{ref}$ and $-\cos\phi_{ref}$ are obtained as 
\begin{eqnarray}
\label{eq:alphabeta}
\left( \begin{array}{c} \sin\phi_{ref} \\ -\cos\phi_{ref} \\ \end{array} \right) = & \sqrt{\frac{2}{3}}\frac{1}{V_{1st, rms}}\left(\begin{array}{ccc} 1 & -\frac{1}{2} & -\frac{1}{2} \\ 0 & \frac{\sqrt{3}}{2} & -\frac{\sqrt{3}}{2} \\  \end{array} \right) \nonumber \\ 
& \times \left( \begin{array}{c} v_{u} \\ v_{v}\\ v_{w} \\ \end{array} \right)
\end{eqnarray}
, where the $V_{1st, rms}$ is the rms voltage of the main grid. Using the measured phase $\phi$, we can extract the deviation of the $\phi$ from the reference $\phi_{ref}$ as 
\begin{eqnarray}
\label{eq:kahou}
\Delta\phi = \phi_{ref}-\phi & \sim  & \sin(\phi_{ref}-\phi) \nonumber \\
& = &\sin\phi_{ref}\cos\phi - \cos\phi_{ref}\sin\phi.
\end{eqnarray}
For the calculation of sine and cosine functions, the algorithm called CORDIC\cite{volder1959cordic} is adopted. Although CORDIC requires only addition, subtraction, bitshift and table lookup, it involves iterative calculations. Therefore, the sampling period 
is changed from 1~$\mu$s to 20~ns only for CORDIC so that many iterations can be done 
within the sampling period of the feedback (1~$\mu$s). In our operation, the number of iterations is set to 11. The PI operator is applied on the $\Delta\phi$ and the result modulates the increment of the counter which corresponds to the measured phase $\phi$.

\subsection{Chopper}
\begin{figure}[!t]
  \centering
  \includegraphics[width=7cm]{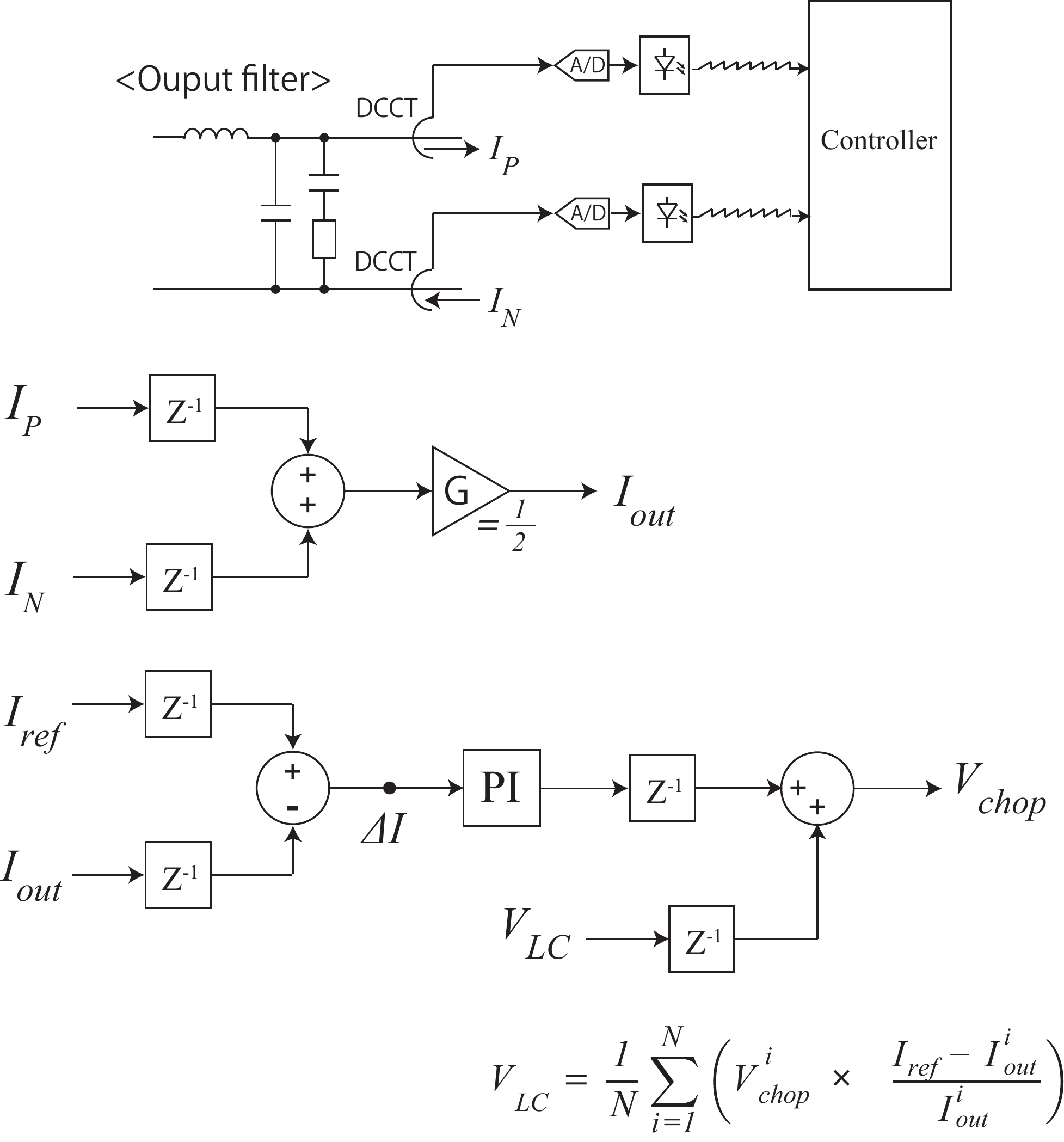}
  \caption{Diagram for chopper control.}
  \label{fig:chop_control_method}
\end{figure}

~The chopper control has been designed based on combination of the
output current feedback with the PI control and the voltage
feedforward, as shown in Figure~\ref{fig:chop_control_method}.
The output current for the feedback (I$_{\rm{out}}$) is obtained by averaging two measured currents through two output terminals (I$_{\rm{P}}$, I$_{\rm{N}}$) so that the common-mode component can be canceled. The current deviation ($\Delta$I), which is the subtraction of the current reference (I$_{\rm{ref}}$) from I$_{\rm{out}}$, is used as the input of the PI controller. Adding the output of the PI controller to the voltage feedforward reference (V$_{\rm{LC}}$) yields the input of the choppers (V$_{\rm{chop}}$) 
The V$_{\rm{LC}}$ can be iteratively modified until the measured current shape becomes same as its reference pattern within the noise level. The details of this iterative learning control is already reported in our previous work \cite{MRMAG2014precise}. At the final stage of the controller, normalizing V$_{\rm{chop}}$ by the measured DC voltage (V$_{\rm{DC}}$ in Figure~\ref{fig:VdcCont}) gives the IGBT duty cycle, which is used as the input of PWM. \\
The sampling frequency of the chopper control is 96 kHz which is matched to the update rate of the reference patterns ($I_{\rm{ref}}$, $V_{\rm{LC}}$). For this firmware, the control algorithm is implemented base on 64-bit floating-point.

\section{Experimental result}
\label{sec:Test}
\begin{figure}[!t]
  \centering
  \includegraphics[width=7.5cm]{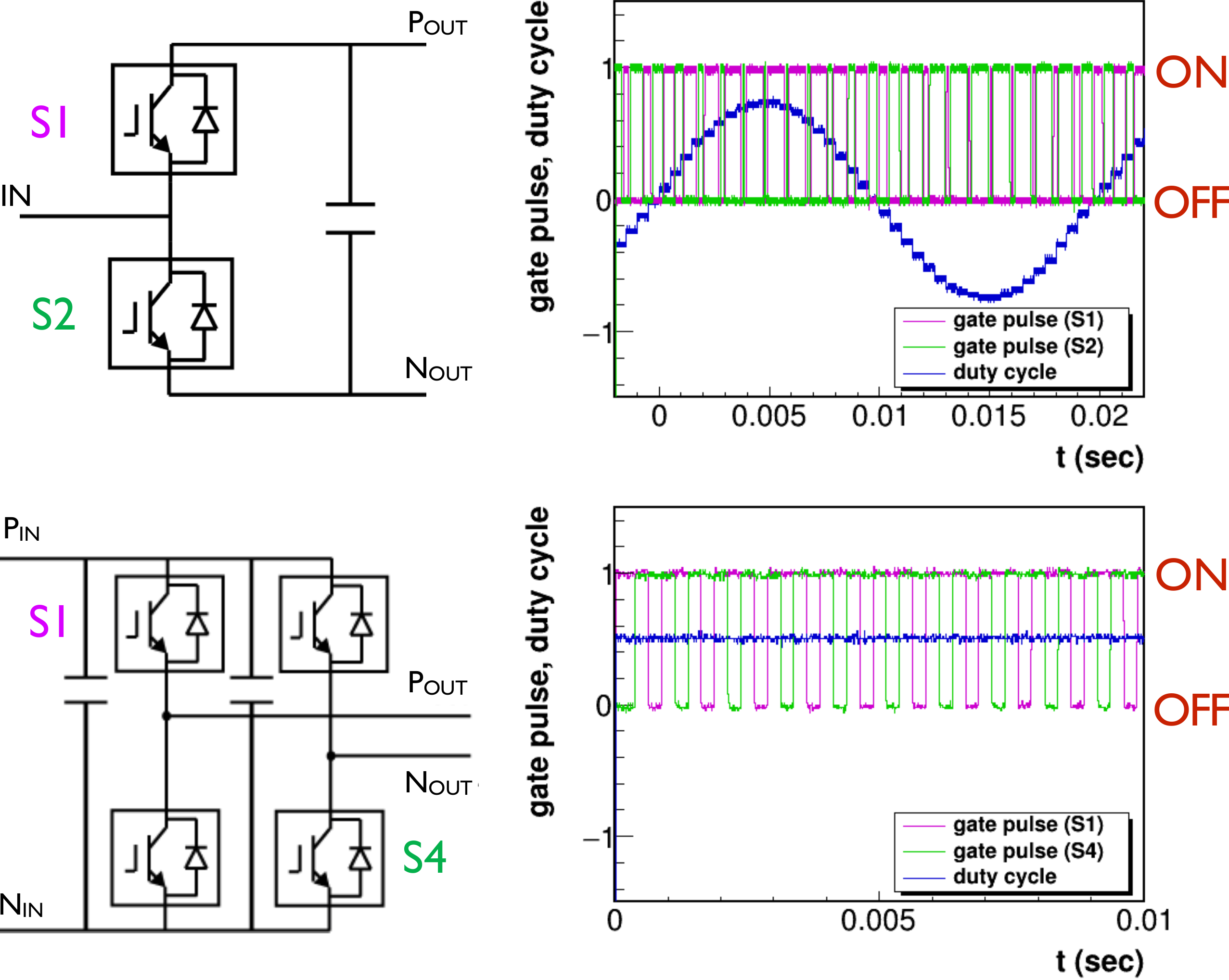}
  \caption{Gate pulses for half-bridge AC/DC converter (top) and full-bridge chopper (bottom).}
  \label{fig:gateCheck}
\end{figure}

The developed control system is combined with a new power converter for small quadrupole magnets and tested. Figure~\ref{fig:gateCheck} shows samples of output gate pulses for a half-bridge AC/DC converter and a full-bridge converters with simulated duty cycles.
The test result of the half-bridge AC/DC converter shows that the gate pulses are modulated by duty cycles. 

\begin{figure}[!t]
  \centering
  \includegraphics[width=7cm]{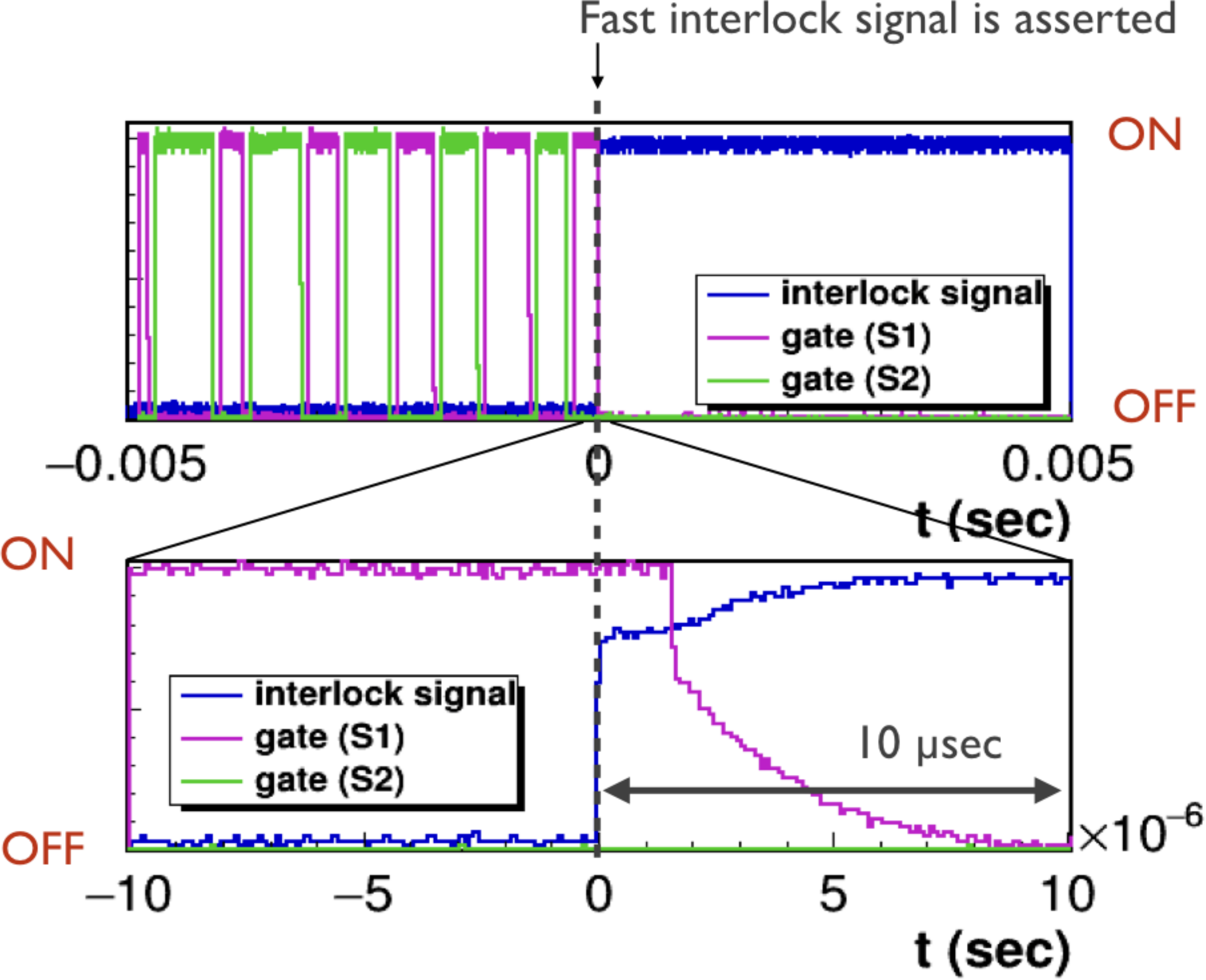}
  \caption{Demonstration of fast interlock system}
  \label{fig:fastInterlock}
\end{figure}

The demonstration of fast interlock system is shown in Figure~\ref{fig:fastInterlock}. We confirmed that the gate pulses of IGBTs are turned off within 10 $\mu$s after a fast interlock signal is asserted. 

\begin{figure}[!t]
  \centering
  \includegraphics[width=7cm]{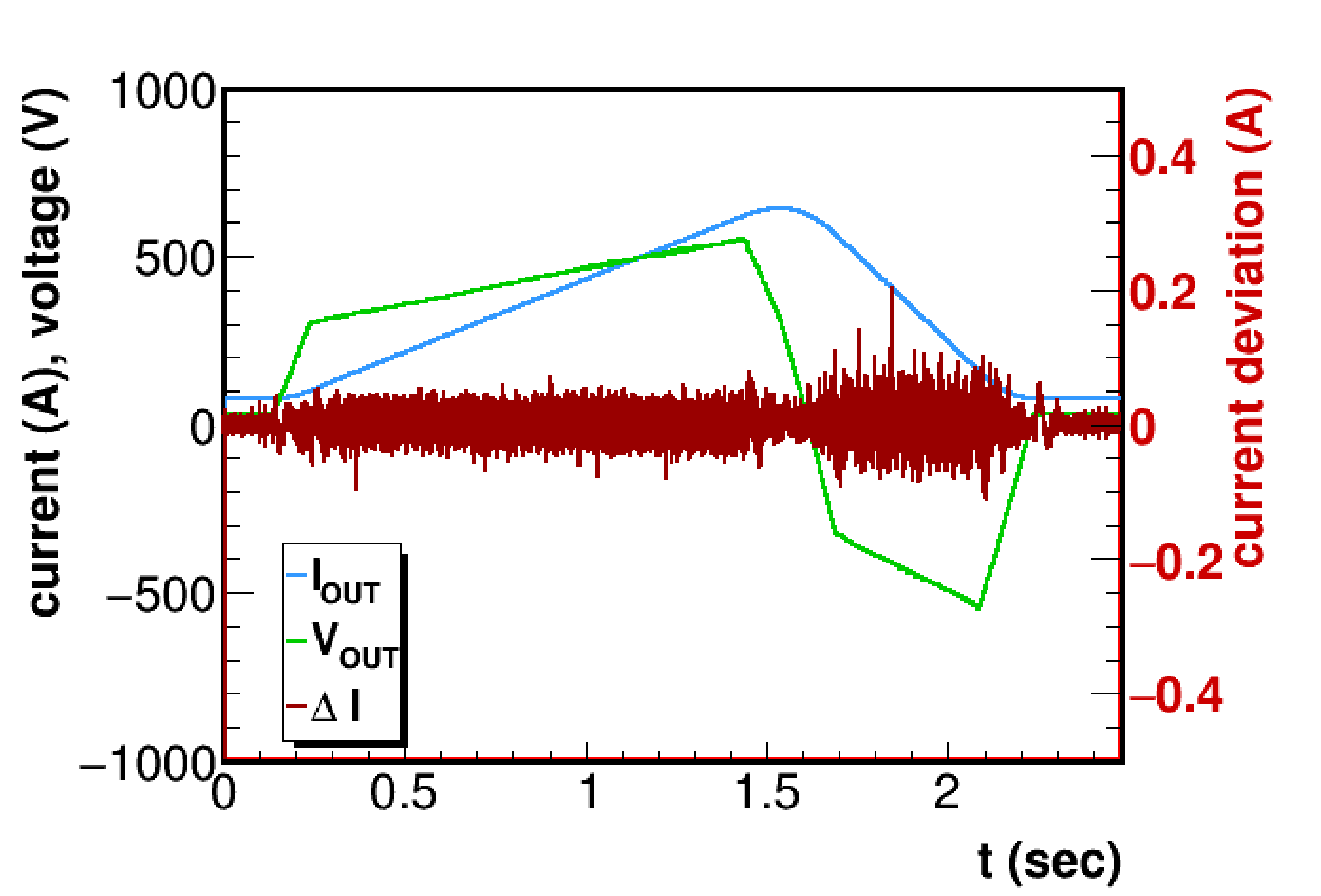}
  \caption{Measured output current, voltage and current deviation with a pattern operation}
  \label{fig:Operation_2480ms}
\end{figure}

Figure~\ref{fig:Operation_2480ms} shows the measured output current, voltage and current deviation for the pattern operation which is a representative current pattern of power converter in J-PARC MR. The maximum and minimum current are set to 75 A and 650 A. The ramping-up period corresponds the present beam acceleration period in J-PARC MR operation. \\

These results show that new control system satisfies our requirement.
In addition, the small quadrupole power converter combined with this control system is operating in present accelerator operation at J-PARC MR.
There are no fatal error in past two years operation.

\section{Conclusion}
A new control system of new power converter for main magnets in J-PARC MR was designed and has been successfully operated with new power converter for small quadrupole magnets with a representative J-PARC MR operation. 
We are ready to apply this control system to new power converters in production. Now, Combined test with a new power converter for the bending magnets is ongoing. 


%

\section*{Acknowledgment}
We would like to thank Mr. Ryu Sagawa, Universal engineering, for this work.

\ifCLASSOPTIONcaptionsoff
  \newpage
\fi

\end{document}